\documentclass
[reprint, showpacs, amsmath, amssymb, aps, floatfix]
{revtex4-1}

\usepackage{xcolor}
\usepackage[utf8]{inputenc}
\usepackage[english]{babel}
\usepackage{graphicx}
\usepackage{comment}
\usepackage{xfrac}

\begin{document}

\title{Synchronization malleability in neural networks under a distance-dependent coupling}
\author{R. C. Budzinski}
\thanks{These authors contributed equally to this paper}
\author{K. L. Rossi}
\thanks{These authors contributed equally to this paper}
\author{B. R. R. Boaretto}
\author{T. L. Prado}
\author{S. R. Lopes}
\email{lopes@fisica.ufpr.br}
\affiliation{Department of Physics, Universidade Federal do Paran\'a, 81531--980, Curitiba, PR, Brazil.}
\begin{abstract}
We investigate the synchronization features of a network of spiking neurons under a distance-dependent coupling following a power-law model. The interplay between topology and coupling strength leads to the existence of different spatiotemporal patterns, corresponding to either non-synchronized or phase-synchronized states. Particularly interesting is what we call synchronization malleability, in which the system depicts significantly different phase synchronization degrees for the same parameters as a consequence of a different ordering of neural inputs. We analyze the functional connectivity of the network by calculating the mutual information between neuronal spike trains, allowing us to characterize the structures of synchronization in the network. We show that these structures are dependent on the ordering of the inputs for the parameter regions where the network presents synchronization malleability and we suggest that this is due to a complex interplay between coupling, connection architecture, and individual neural inputs.
\end{abstract}

\maketitle

\section{Introduction}

Modeling of natural phenomena through the use of coupled networks finds applications in various scientific areas \cite{Strogatz_2001,boccara_2010}. Examples can be found in statistical physics \cite{barthelemy1999small}, power grid distributions \cite{kinney2005modeling}, ecology \cite{banerjee2016chimera}, and neuroscience \cite{bassett2017network}. This approach has revealed a variety of synchronization phenomena in coupled networks, such as complete synchronization, and phase synchronization (PS) \cite{rosenblum1996phase, boccaletti2006complex, arenas2008synchronization}. Building on these findings, recent investigations have shown cases where the network susceptibility plays an important role so that small changes in the system can produce significant consequences \cite{manik2017network}. Examples are in synchronization vulnerability, where small perturbations in nodes can lead to a desynchronization process \cite{medeiros2019state}, and also changes in a small number of initial conditions can lead to different dynamical features related to chimera states \cite{santos2018riddling}.

In neuroscience, synchronization has been found to be of fundamental importance: it has been observed in healthy behaviors, like memory \cite{fell2011role}, conscious processes \cite{Gaillard_consciousProcessing}, visual-motor behavior \cite{roelfsema1997visuomotor} and perception phenomena \cite{rodriguez1999perception}. Also, either excess or lack of synchronization have been related to unhealthy behaviors, like seizures, generating epileptic episodes \cite{mormann2000mean}, Parkinson's disease \cite{galvan2008pathophysiology}, and autism \cite{dinstein2011disrupted}.

One way to study synchronization is by the modeling of a neuronal network. There are several models to reproduce the dynamics observed in neurons \cite{ibarz2011map}. A suitable model is the two-dimensional map developed by Chialvo \cite{chialvo1995generic}, which mimics spiking behavior. Networks composed of neurons simulated by this model can present a diversity of dynamical phenomena, such as bistability with explosive synchronization (in small-world networks) \cite{boaretto2019mechanism} or even the emergence of synchronization patterns like clustering synchronization and anti-phase clusters, occurring due to the interplay of interacting sub-populations \cite{kamal2015emergent}.

Besides the neuronal model, the connection topology is also an important factor for synchronization phenomena \cite{gomez2007paths}. In general, it has been found that connection architectures composed of local structures do not facilitate high degrees of synchronization \cite{kaneko1989pattern}. However, long-range and global connection schemes do facilitate synchronization processes \cite{kaneko1990clustering}. To analyze these phenomena, a coupling architecture given by a distance-dependent power-law scheme can be used. In this kind of coupling, all neurons are connected, but not necessarily all connections are effective since the contribution of more distant nodes decreases with the increase of the power-law exponent. In fact, by varying this parameter, there is a continuous transition from global effectiveness (all neurons contributing equally) to local effectiveness (only first neighbors contributing). This topology has been studied in several contexts since the long-range interaction is observed in fundamental laws of physics, in coupled-oscillators networks \cite{rosenblum2004controlling}, and in biological networks \cite{banerjee2016chimera}. Specifically, in neural systems, a power-law decay was observed in the weight-distance relationship of the mouse connectome \cite{rubinov2015wiring, choi2019synchronization, Knox2018}.

In this paper, we focus on the synchronization properties of networks of Chialvo neurons coupled through the distance-dependent power-law scheme. We use the Kuramoto order parameter \cite{kuramoto2012chemical} to measure PS, with the neuronal spiking activities being associated with geometric phases. The main results consist of the description and analysis of the phenomenon we call synchronization malleability, in which the PS behavior of the network drastically changes as a consequence of variations of the neuronal inputs. We consider a sequence of neuronal input values with a uniform distribution and show that a shuffling process over this sequence can change the network from a highly phase-synchronized state to a highly desynchronized one. Besides, we show that these systems present diverse synchronization patterns due to the interplay between the coupling strength and the power-law exponent. For networks where the dependence on distance is weaker (closer to the global case), we observe a traditional transition from non-synchronized to phase-synchronized states as the coupling strength is increased. On the other hand, as we make the distant-dependence stronger (closer to the local case), there appear states where only parts of the network are phase-synchronized and there appear also diagonal spatiotemporal structures, or zig-zag states \cite{wang2008synchronization, osipov2005synchronized}. For sufficiently strong distance-dependence, these states start to dominate and the network no longer displays transitions to phase synchronization. 

We also measure the mutual information \cite{cover1991elements} shared between neurons, evaluated through their spike-count \cite{rieke1999spikes, martinoia2009}, which is a measure of their PS degree \cite{palus1997detecting}. With this, we obtain the functional connectivity of the network, allowing us to characterize its synchronization structures. When the information is effectively shared between neurons at both local (neighborhood) and global (long-range) levels, network PS is reached. Otherwise, in the cases where the network is non-phase-synchronized the sharing of information is effective only at the local level. We see that different shuffles of the neuronal inputs may facilitate the sharing of information at the global level, leading to network PS. However, they may also hinder it, resulting in a PS degree smaller than observed in the uncoupled case. 

The paper is organized as follows: in section \ref{sec:2}, the local dynamics and the connection architecture are presented; in section \ref{sec:3}, the methodology is exposed; in section \ref{sec:4}, the main results are shown and the discussions are presented to support the conclusions in section \ref{sec:5}.

\section{The model and connection architecture}\label{sec:2}

To simulate the local dynamics, we consider the neuronal model \cite{chialvo1995generic}
\begin{align}
x_{i,t+1} &= x_{i,t}^2 \exp(y_{i,t} - x_{i,t}) + K_i + I_{i,t}, \label{chialvo1} \\
y_{i,t+1} &= a y_{i,t} - b x_{i,t} + c,\label{chialvo2}
\end{align}
where $x_{i,t}$ and $y_{i,t}$ are the activation and recovery variables of the $i$-th neuron, with the variable $x_{i,t}$ mimicking the membrane potential. $K_i$ is the input signal in each neuron, which acts as an additive perturbation \cite{chialvo1995generic}, affecting the neuron's firing rate. The constants $a$, $b $, and $c$ are parameters that control the dynamical behavior of the model, set to obtain the spiking behavior ($a = 0.89$, $b = 0.6$, and $c = 0.28$) \cite{chialvo1995generic}. $I_{i,t}$ is the coupling term between neurons and follows a distance-dependent power-law scheme described as
\begin{equation}
I_{i,t}=\frac{\varepsilon}{\eta^{\alpha}} \sum \limits_{j=1}^{N'} \frac{x_{i-j,t}+x_{i+j,t}}{j^{\alpha}},
\label{coupling}
\end{equation}
where $\varepsilon$ is the coupling strength, $N$ is the network size, $N' = (N-1)/2$, $\alpha$ is the power-law exponent, or locality parameter, and $\eta$ is the normalization factor given by 
\begin{equation}
\eta^{\alpha}=2\sum\limits_{j=1}^{N'}\frac{1}{j^{\alpha}}.
\end{equation}
It is important to notice that $\alpha$ controls the range of effective connections, which contribute significantly to the coupling term $I_{i,t}$. When $\alpha = 0$, a typical global coupling scheme is obtained, where the $i$-th neuron is effectively coupled with all the other $N-1$ neurons. When $\alpha$ is increased, the distance-dependence becomes stronger, and more distant neurons contribute less. In the extreme case of $\alpha \rightarrow \infty$, the coupling scheme is characterized by a first-neighborhood topology. In all cases, the network follows ring topology, i.e. a one-dimensional network with periodic boundaries. 

\begin{figure}[htb]
    \centering
    \includegraphics[width=0.9\columnwidth]{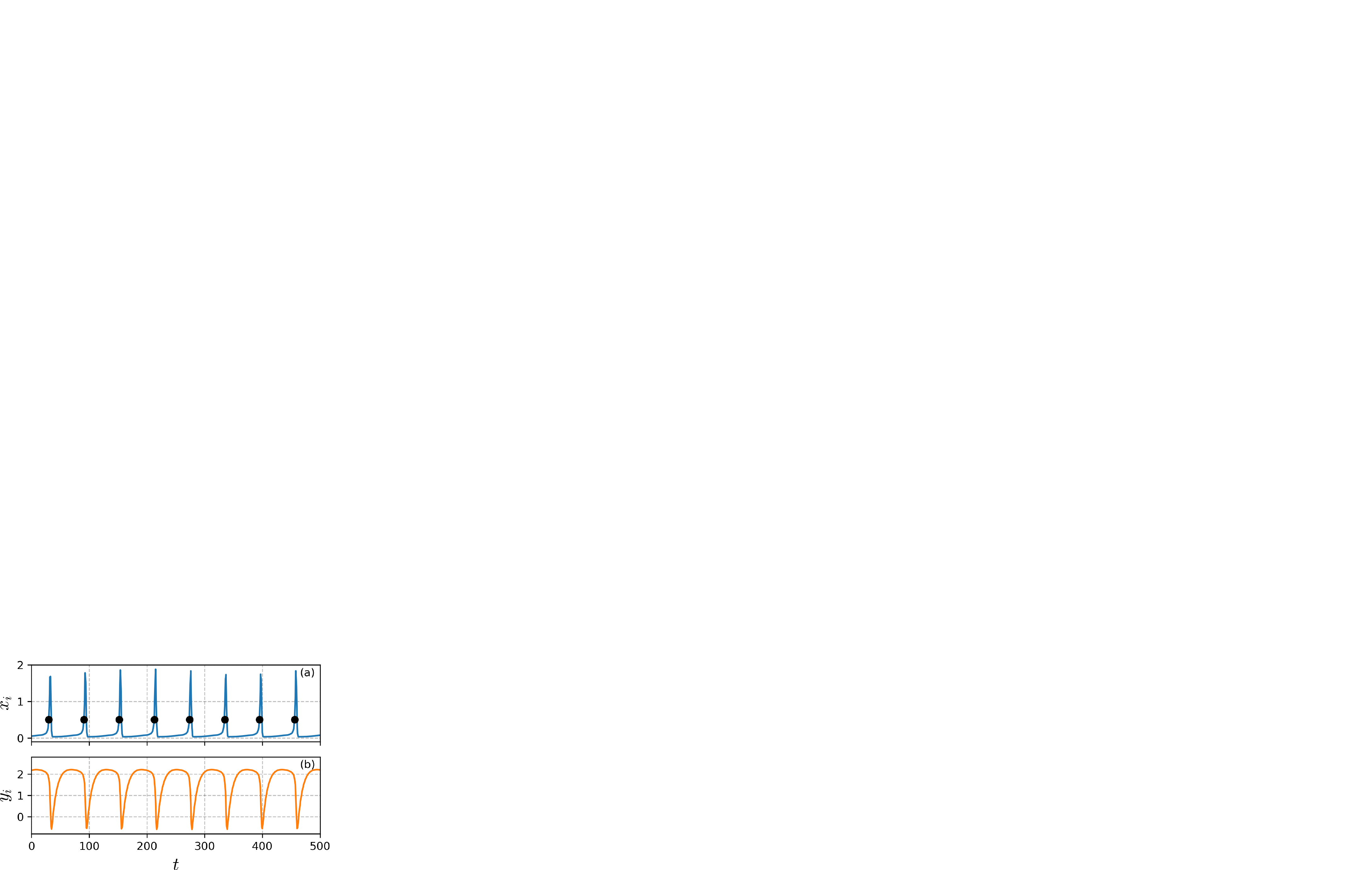}
    \caption{The dynamics of an isolated neuron following Eqs. (\ref{chialvo1}) and (\ref{chialvo2}), with $K_i=0.03$. Panel (a) shows the activation variable (membrane potential), and panel (b) depicts the recovery variable. The black circles denote the spike times.}
    \label{fig:dynamics}
\end{figure}

Figure \ref{fig:dynamics} depicts the dynamics of one isolated neuron. Panel (a) shows $x_{i,t}$ as a function of $t$, characterizing a depolarization and repolarization process (spiking dynamics). Panel (b) depicts the recovery variable dynamics $y_{i,t}$. For the entire paper, similar dynamical features are observed for all coupled neurons. The black circles indicate when the activation variable reaches the condition $x_{i} = 0.5$ with positive first derivative, leading to a spike event. The sequence of spike times for one neuron is called its spike train, which we use in the evaluation of phases and production of raster plots.

The input values $K_{i}$ are initially built to produce a uniform distribution
\begin{equation}
K_{i} = 0.03 + \frac{i\sigma}{N},
\label{eq:inputs}
\end{equation}
where $\sigma = 0.0035$ is the coefficient of neuronal dissimilitude. Different values of $\sigma$ were tested and similar results were obtained, as long as $\sigma$ is kept sufficiently small to guarantee the spiking behavior \cite{chialvo1995generic}.

For the simulations, we perform a random shuffling process over the $K_i$ values obtained from Eq. (\ref{eq:inputs}). This leads to a different ordering in the input values but keeps the distribution still uniform. We consider $30$ different shuffled sequences, labeled as shuffling \#1, shuffling \#2, ..., shuffling \#30. The relevant codes and sequences of input values ($\{K_{i}\}$) can be found in the repository \cite{git_hub_kalel}. Similar results are also observed when $K_{i}$ are obtained from random generators considering the limits $[0.03,\,0.03+\sigma]$.

\begin{figure}[htb]
    \centering
    \includegraphics[width=0.95\columnwidth]{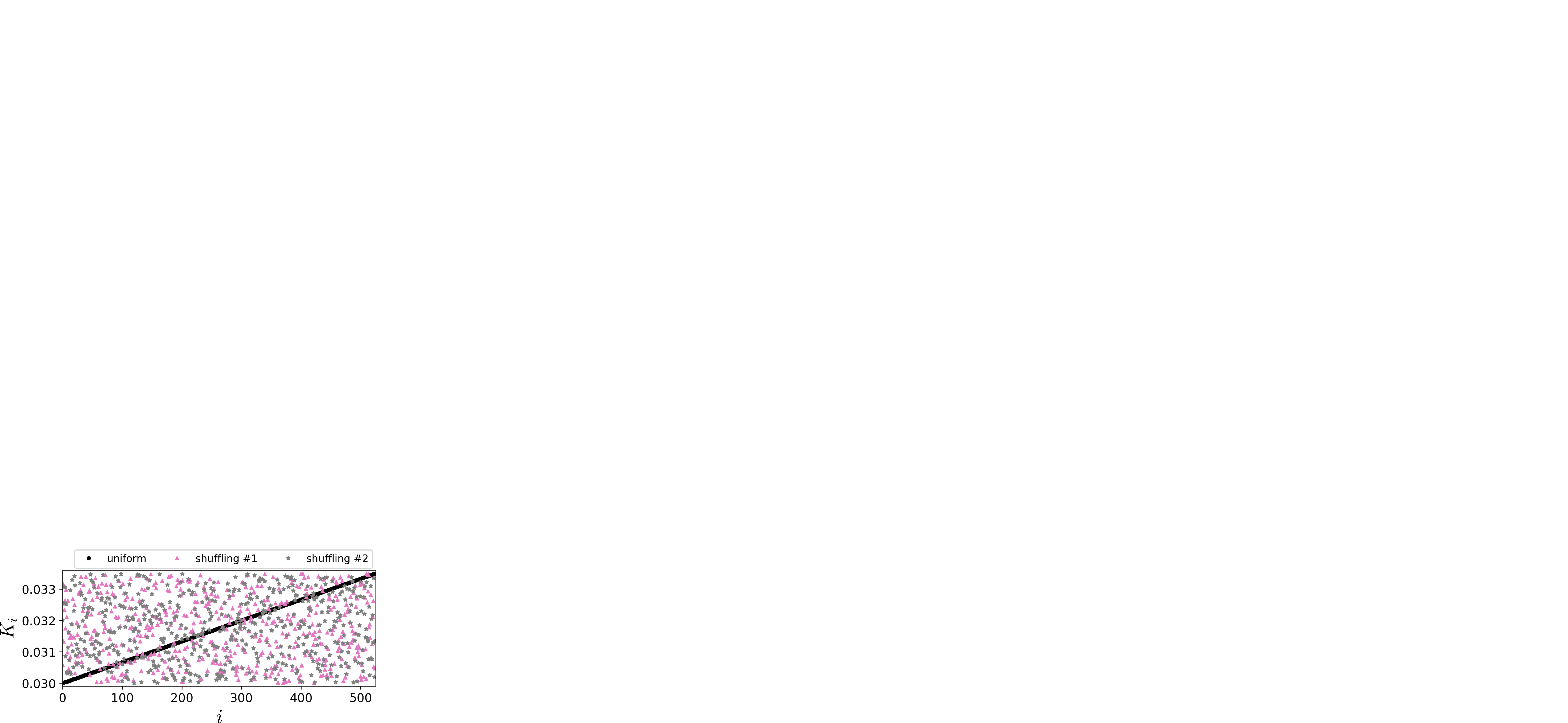}
    \caption{Input values $K_{i}$ as a function of neurons' index $i$. The black dots represent the case where $K_{i}$ is built following Eq. (\ref{eq:inputs}) and pink up triangles and gray stars represent two shuffled cases labeled as shuffling \#1 and shuffling \#2.}
    \label{fig:inputs}
\end{figure}

Figure \ref{fig:inputs} shows the values of the inputs $K_{i}$ as a function of the neurons' index $i$. The black dots represent the case constructed following Eq. (\ref{eq:inputs}). The pink up triangles and the gray stars correspond to shuffling \#1 and \#2, respectively. 

\section{Quantifiers}\label{sec:3}

\subsection{Kuramoto order parameter}\label{sec:3.1}

The Kuramoto order parameter \cite{kuramoto2012chemical} is used to quantify phase synchronization (PS) between oscillators. To do that, we first define a phase $\theta_{i}$ for the $i$-th neuron such that $\theta_i$ increases by $2\pi$ for every spike. A continuous variation of $\theta_{i}$ can be obtained through a linear interpolation \cite{boccaletti_2002}
\begin{equation}
\theta_{i}(t) = 2\pi n_i + 2\pi \frac{t - t_{n,i}}{t_{n+1, i} - t_{n,i}}, \; (t_{n,i} \leq t < t_{n+1,i}),
\label{eq:phase}
\end{equation}
where $t_{n,i}$ is the time when the $n$-th spike occurs in the $i$-th neuron.

The degree of PS of the network for each time $t$ is, then,
\begin{equation}
\label{eq:kuramoto}
R(t) = \frac{1}{N}\left|\sum\limits_{j=1}^N \exp{(i \theta_j(t))}\right|,
\end{equation}
with $i$ in this equation denoting the imaginary unit. The quantifier $R$ ranges between $0$ and $1$. If $R = 1$ there is complete phase synchronization. Otherwise, if $R \rightarrow 0$ the system can be non-synchronized, meaning spatiotemporal incoherence, or it can even display anti-phase-synchronization.

We take the time average to obtain the mean Kuramoto order parameter $\langle R \rangle$:
\begin{equation}
\label{eq:kuramoto_mean}
    \langle R \rangle = \frac{1}{t_{\mathrm{f}} - t_{0}} \sum_{t=t_0}^{t_\mathrm{f}} R(t),
\end{equation}
in which $t_{0}$ is the transient time and $t_{\mathrm{f}}$ is the whole simulation duration.

The Kuramoto order parameter can also be evaluated for only a group of neurons in the network \cite{shanahan2010metastable}:
\begin{equation}
    \langle R^{j} \rangle = \frac{1}{t_{\mathrm{f}} - t_{0}}\sum\limits_{t = t_{0}}^{t_{\mathrm{f}}} \frac{1}{N_{\mathrm{local}}} \left|\sum\limits_{k\in\Omega_j} \exp{(i \theta_k(t))}\right|,
    \label{eq:mean_kuramoto_local}
\end{equation}
where the phases $\theta_{k}(t)$ are obtained from Eq. (\ref{eq:phase}), $N_{\mathrm{local}} = N/M$ is the number of neurons in each group when the entire network is divided into $M$ groups, and $\Omega_{j}$ is the set of neurons belonging to the $j$-th group. The partitioning is done such that the $j$-th group $\Omega_{j}$ contains neurons with indices in the interval $[j N_{\mathrm{local}}, (j+1) N_{\mathrm{local}})$, with $j = 0,1, \cdots, M-1$.

Eq. (\ref{eq:mean_kuramoto_local}) allows us to compare the PS in the groups and the entire network by defining
\begin{equation}
    \delta R = \overline{\langle R_{\mathrm{local}} \rangle} - \langle R \rangle,
    \label{eq:deltaR}
\end{equation}
where $\overline{\langle R_{\mathrm{local}} \rangle} = \sfrac{1}{M} \sum_{j=1}^{M} \langle R^{j} \rangle$ is the average over all $M$ groups. If $\delta R \rightarrow 1$, the groups of neurons are phase synchronized but the entire network is not. On the other hand, if $\delta R \rightarrow 0$, the groups and the entire network have similar dynamics, either non-synchronized or phase synchronized.

\subsection{Spiking frequency}\label{sec:3.2}

Given the phases $\theta_{i}(t)$ associated to each neuron in the network, we evaluate the angular frequency related to the spiking activity as:
\begin{equation}
\omega_{i} = \frac{d\theta_{i}(t)}{dt} = \lim_{t \to \infty} \frac{\theta_{i}(t)-\theta_{i}(0)}{t}.
\label{eq:spiking_freq}
\end{equation}

In this way, to quantify the degree of frequency synchronization in the network, we evaluate the standard deviation of the spiking frequency $\omega_{i}$ over all neurons in the network and then divide by its mean value:
\begin{equation}
\kappa(\omega) = \frac{\sqrt{\frac{\sum\limits_{i=1}^{N}\left(\omega_{i} - \overline{\omega}\right)^{2}}{N}}}{\overline{\omega}},
\label{eq:sinc_freq}
\end{equation}
where $\overline{\omega} = \sfrac{1}{N} \sum_{i=1}^{N} \omega_{i}$ is the mean frequency. In this sense, if $\kappa(\omega) \rightarrow 0$, neurons have the same spiking frequency and the network is in a state with frequency synchronization.   

\subsection{Information analysis}\label{sec:3.3}

In this subsection, we describe the quantifiers we used to measure the information contained in the neuronal spike trains. The coding mechanism we consider is the spike-count code, according to which information is coded by the neuron in the number of spikes over some relevant time window \cite{adrian1928}. In this case, we partition the spike train into bins of size $\Delta t$ and count the number of spikes per bin \cite{rieke1999spikes, martinoia2009}. The choice of the parameter $\Delta t$ is discussed later, at the end of this subsection. Then, the information in the spike train of neuron $\mathcal{N}$ is calculated as its Shannon entropy \cite{shannon1948theoryofcommunication}:
\begin{equation}
    H(\mathcal{N}) = -\sum_{n \in \mathcal{N}} p(n) \log_2{p(n)},
\end{equation}
where $n$ is the number of spikes in each bin and $p(n)$ is the probability of observing $n$ spikes, estimated as the frequency of bins with $n$ spikes in the time series.

For two neurons $\mathcal{M}$ and $\mathcal{N}$, we can calculate the amount of information shared by these two (or, the amount of information that one contains about the other) using the mutual information \cite{cover1991elements, deschle2019directed}:
\begin{equation}
    MI(\mathcal{M},\mathcal{N}) = \sum_{m \in \mathcal{M}}\sum_{n \in \mathcal{N}} p(m,n) \log_2 \frac{p(m,n)}{p(m)p(n)},
\end{equation}
where $p(m)$ and $p(n)$ are marginal probabilities and $p(m,n)$ is the joint probability of the values $m$ and $n$. For example, $p(1,2)$ is the frequency of appearance of $1$ spike in neuron $\mathcal{M}$ and $2$ spikes in neuron $\mathcal{N}$ in the same bin throughout the series. The mutual information is symmetric, $MI(\mathcal{M},\mathcal{N}) = MI(\mathcal{N},\mathcal{M})$, and reduces to the entropy $H(\mathcal{M})$ when $\mathcal{M} = \mathcal{N}$. Mutual information is a very useful tool for measuring statistical correlations \cite{martinoia2009}, being capable of detecting nonlinear interactions \cite{Timme2018, quiroga2013principles}.

Furthermore, this method of calculating MI between two neurons is also a measure of their PS \cite{palus1997detecting}. If two neurons $\mathcal{M}$ and $\mathcal{N}$ are completely phase desynchronized, their spike trains are completely uncorrelated, so $MI(\mathcal{M},\mathcal{N}) = 0$. If both are completely phase synchronized, their spike trains are the same, so $MI(\mathcal{M},\mathcal{N}) = H(\mathcal{M}) = H(\mathcal{N})$. This enables us to study the amount of shared information between the neurons and the phase synchronization patterns of the network.

For a wide range of $\Delta t$, an oscillating pattern in the mutual information between neurons is observed. We choose the $\Delta t$ that corresponds to the first maximum, following the principle of maximization of entropy \cite{rieke1999spikes} and giving us the clearest results. Other values lead to qualitatively similar results, as the patterns observed remain the same. The only difference is quantitative, in the absolute values of the $MI$. This was verified through several simulations using different values of $\Delta t$. As a consequence of our choice, the bin size $\Delta t$ isn't necessarily the same for every parameter value. 

The challenge with using $MI$ is the estimation of the probability distributions. We use the method known as the plug-in, or direct method \cite{Borst1999, quiroga2013principles}, which results in positively biased estimates. However, this positive bias can be reduced by increasing the number of samples \cite{quiroga2013principles, treves_panzeri_1995}. We consider $500000$ times for the analyses, which give us stable results: increasing this time did not alter results considerably.

At last, we collect the mutual information between every pair of neurons $(\mathcal{M},\mathcal{N})$ in a matrix $(\mathbf{MI})_{\mathcal{M}\mathcal{N}} = MI(\mathcal{M},\mathcal{N})$ and consider this as the functional connectivity (FC) of the network.

\section{Results and discussion}\label{sec:4}

We study a network composed of $N = 525$ spiking neurons following Eqs. (\ref{chialvo1}), (\ref{chialvo2}), and (\ref{coupling}), where $\alpha \in [0.0,\,3.5]$ (locality parameter) and $\varepsilon \in [0.000,0.099]$ (coupling strength). We consider $30$ simulations, each one with a different shuffled sequence of $K_{i}$ values. The initial conditions for the $N$ neurons are randomly chosen in the intervals $x \in [0.0,\,2.0]$ and $y \in [-1.0,\,2.0]$, but are kept the same for different simulations. The transient time is given by $t_{0} = 100000$ times and the total simulation time is $t_{\mathrm{f}} = 200000$. The dynamical features of the systems are completely captured considering this simulation time, but in the mutual information analysis, we consider $t_{\mathrm{f}} = 500000$ in order to obtain more data and ensure that the mutual information estimates are robust. The analyses on the Kuramoto order parameter for the groups is performed for $M = 15$, which leads to $N_{\mathrm{local}} = 35$ neurons in each group.

\subsection{Synchronization scenario}
\begin{figure*}[htb]
    \centering
    \includegraphics[width=0.95\textwidth]{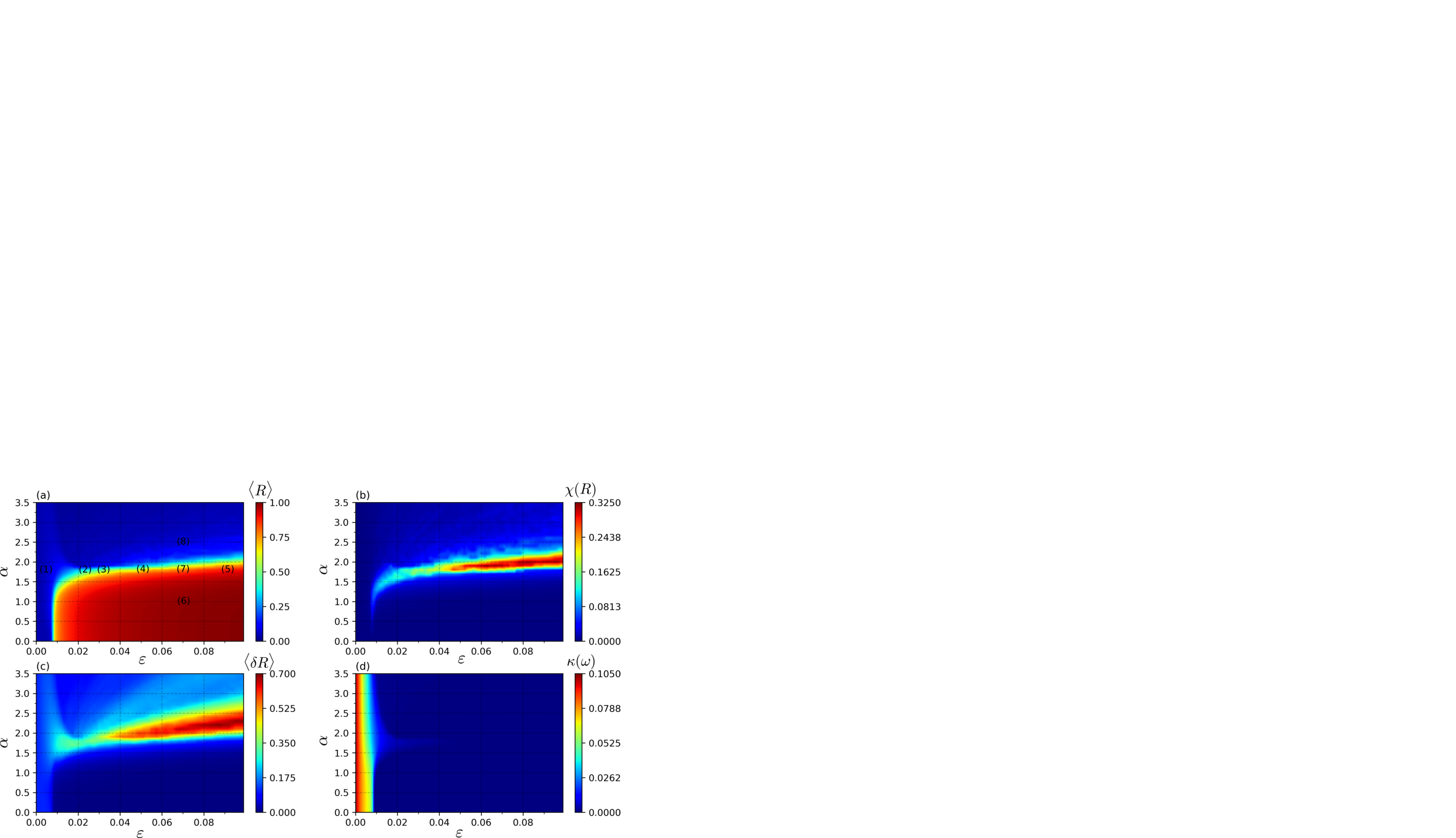}
    \caption{The synchronization scenario in the parameter space $\alpha \times \varepsilon$. Results are an average taken over $30$ simulations with different sequences of inputs $K_{i}$ and fixed initial conditions for ($x_{i}$,  $y_{i})$. Panel (a) depicts the average degree of phase synchronization (PS) $\langle R \rangle$ where one can observed PS (red tones) and non-phased-synchronized states (blue tones). The numbers indicate regions for future reference. Panel (b) shows the standard deviation of $\langle R \rangle$ ($\chi(R)$) over the $30$ simulations, where higher values (red tones) indicate that the system can present different values of $\langle R \rangle$ in each simulation. Panel (c) depicts $\langle \delta R \rangle$, where higher values indicate that the network has phase-synchronized groups, but is phase-desynchronized as a whole. Panel (d) depicts $\kappa (\omega)$, where smaller values (blue tones) denote regions where the networks are frequency-synchronized.}
    \label{fig:main_sync}
\end{figure*}

Figure \ref{fig:main_sync} depicts synchronization features given by an average over $30$ simulations in the parameter space of $\alpha \times \varepsilon$. Panel (a) depicts the average degree of PS in the heatmap of the temporal average of the Kuramoto order parameter for the entire network ($\langle R \rangle$). For all values of $\alpha$ shown, with $\varepsilon \lesssim 0.01$, the network is non-phase-synchronized (dark blue tones). However, for $\alpha \lesssim 1.5$, the increase of coupling strength $\varepsilon$ leads the network to phase-synchronized states (red tones), a traditional scenario observed in several contexts \cite{gomez2007paths,kuramoto2012chemical,boccaletti_2002}. Increasing the locality parameter to the range $1.5 \lesssim \alpha \lesssim 2.0$, the transition to PS is maintained, but on average with a reduced degree (smaller $\langle R \rangle$) and a higher $\varepsilon$ value being required to reach phase-synchronized states. At last, for $\alpha \gtrsim 2.0$, the network no longer transitions to PS, since, for the entire interval of $\varepsilon$, $\langle R \rangle$ depicts blue tones. In this case, the increase of the coupling strength ($\varepsilon \gtrsim 0.01$) actually reduces the average degree of PS and the network depicts values of $\langle R \rangle$ smaller than the uncoupled case. Hence, we observe two distinct transitions between non-phase-synchronized states and phase-synchronized states: one induced by the increase of coupling strength and another by the decrease of $\alpha$. In the former, an increase of $\varepsilon$ leads to PS. In the latter, an increase of $\alpha$ terminates PS, as the network becomes locally coupled. A similar scenario was found considering different local dynamics \cite{anteneodo2003analytical, marodi2002synchronization, li2006phase}.

An interesting behavior is observed in panel (b) of Fig. \ref{fig:main_sync}, where the dispersion of the networks' PS degrees is shown. In this case, higher values of $\chi(R)$ indicate that the system can depict different degrees of PS in different simulations. This is most pronounced in $1.6 \lesssim \alpha \lesssim 2.3$ and $\varepsilon \gtrsim 0.020$, where there is the transition to PS induced by a decrease in $\alpha$ (as seen in panel (a)). This region is called malleable, since the shuffling process in the input values $K_{i}$ leads to networks presenting different degrees of PS, and, consequently, different dynamical states for the same set of parameter $\alpha$ and $\varepsilon$. For other regions in the parameter space ($\varepsilon \lesssim 0.020$ or $\alpha \gtrsim 2.3$ or $\alpha \lesssim 1.6$), the quantifier $\chi(R)$ depicts small values, indicating that the network's dynamics is similar over simulations. We note that the synchronization malleability phenomenon occurs for intermediate values of $\alpha$, between the extreme cases of local and global effectiveness. A similar phenomenon was observed in a small-world network of bursting neurons when some local connections are changed by non-local ones \cite{budzinski2019synchronous}.

Panel (c) depicts the quantifier $\langle \delta R\rangle$, given by Eq. (\ref{eq:mean_kuramoto_local}). For $\varepsilon \gtrsim 0.020$ and $1.5 \lesssim \alpha \lesssim 2.7$ higher values are observed, indicating that there is PS in the groups of neurons, but not in the entire network. Particularly interesting is the case for $1.6 \lesssim \alpha \lesssim 2.0$, since the increase of $\varepsilon$, for a fixed value of $\alpha$, makes $\langle \delta R \rangle$ transition from small (blue tones) to high values (red tones) and then back to small values (blue tones). For these regions, the same increase of $\varepsilon$ makes the network transition from non-phase-synchronized states to phase-synchronized ones (see panel (a)), where $\delta R$ indicates that this transition occurs with the existence of PS at the group level before the entire network reaches phase-synchronized states. For $\alpha \lesssim 1.5$, $\langle \delta R\rangle$ always depicts small values, and the transition to PS induced by the coupling strength does not pass through states with group PS. For high values of $\alpha$, $\alpha \gtrsim 2.9$, the quantifier shows also small values, since the network does not reach PS. These analyses were performed considering $M = 15$ groups, but similar results are obtained considering $M = 7$ and $M = 5$. The absolute value changes, but the region in the parameter space where the quantifier depicts higher values is the same, which corroborates the main idea of the existence of phase-synchronization on a local level, but not on a global one. Details of the dynamical states can be observed in the raster plots (RP) of the network (see Figs. \ref{fig:raster_plot_1} and \ref{fig:raster_plot_2}).

At last, panel (d) of Fig. \ref{fig:main_sync} shows the quantifier $\kappa (\omega)$, described by Eq. (\ref{eq:sinc_freq}). For the entire interval of $\alpha$ and $\varepsilon \lesssim 0.010$, higher values indicate that the network is not frequency synchronized. The increase of the coupling strength beyond $\varepsilon \gtrsim 0.010$ then leads the network to frequency-synchronized states. This phenomenon is observed for small values of $\alpha$, where the network is on a phase-synchronized state (high value of $\langle R \rangle$), which is expected. However, for higher values of $\alpha$, the network is also frequency synchronized, but not phase synchronized.

\subsection{Malleability analysis}
\begin{figure*}[ht]
    \centering
    \includegraphics[width=0.85\textwidth]{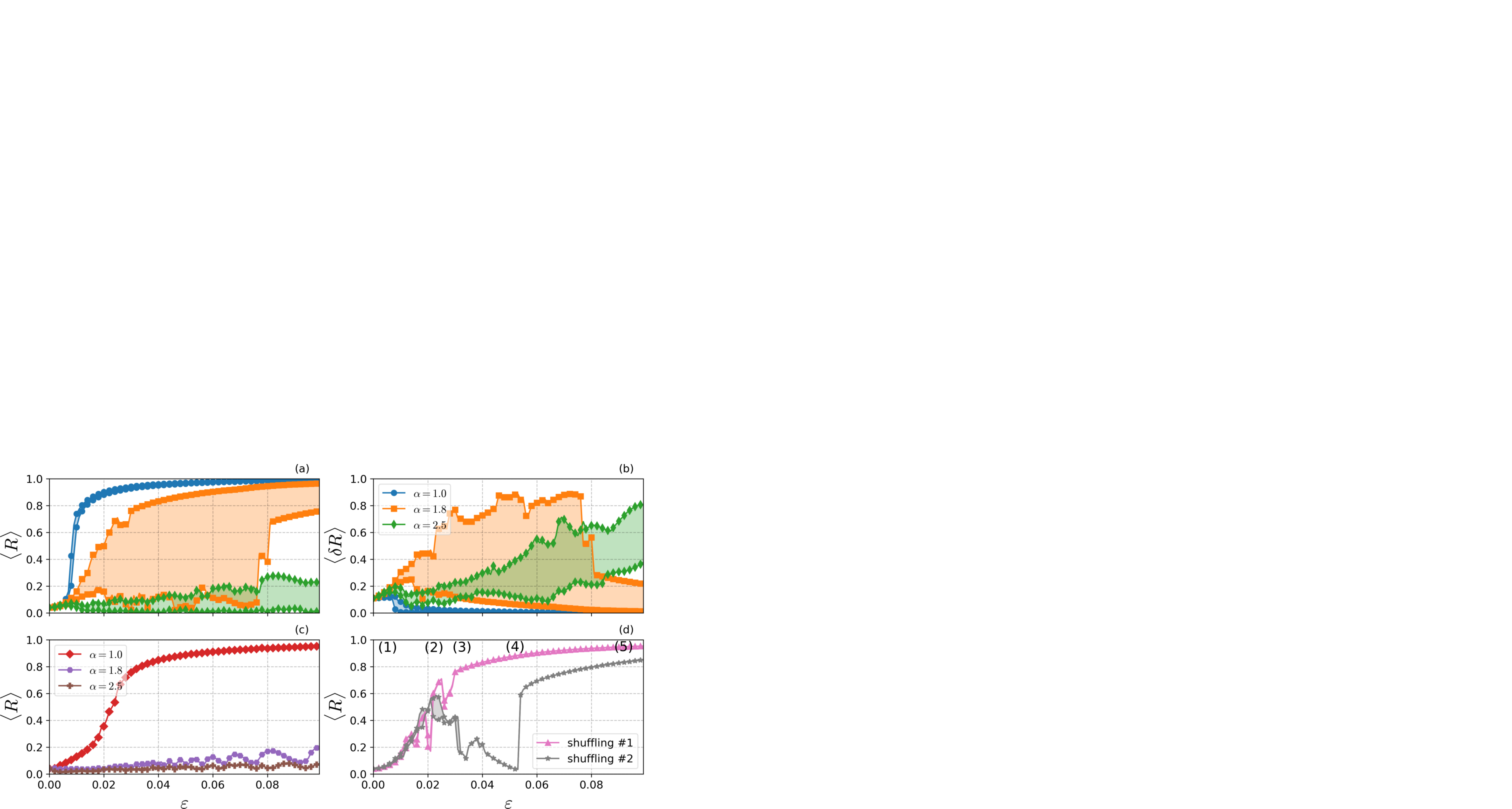}
    \caption{Panels (a) and (b) depict the maximum and minimum values (extremes of the filled area) of $\langle R \rangle$ and $\langle \delta R \rangle$, respectively, as a function of $\varepsilon$. This is done considering $30$ simulations with different shuffles over the inputs values ($K_{i}$) and fixed initial conditions for $x_{i}$ and $y_{i}$. Moreover, panel (c) depicts the temporal average of the Kuramoto order parameter as a function of $\varepsilon$ for the network where the input values are not shuffled (Eq. (\ref{eq:inputs})). At last, panel (d) depicts $\langle R \rangle$ as a function of $\varepsilon$ for two cases with fixed shuffled input values, shuffling \#1 (pink line with up triangles) and shuffling \#2 (gray line with starts) (these cases are represented in Fig. \ref{fig:inputs}), considering $\alpha = 1.8$ for both. Here, $30$ simulations with different initial conditions for $x_{i}$ and $y_{i}$ are considered and only the maximum and minimum values are shown.}
    \label{fig:order_cases}
\end{figure*}

In order to analyze the details of the network's dynamics, Fig. \ref{fig:order_cases} (a) depicts the maximum and minimum values of $\langle R \rangle$, represented by the extremes of the filled area, for the $30$ simulations as a function of the coupling strength $\varepsilon$. The same representation is used in panel (b) for the maximum and minimum values of $\langle \delta R \rangle$. For $\alpha = 1.0$ (blue lines with circles), an increase in $\varepsilon$ makes the network reach PS, as $\langle R \rangle$ increases  and $\langle \delta R \rangle$ is small for all simulations. This is expected since in Fig. \ref{fig:main_sync} (b) $\chi(R)$ shows a small value in this region. 

On the other hand, for $\alpha = 1.8$ (orange lines with squares), the network may assume significantly different synchronization states, as seen in the difference between the extremes of $\langle R \rangle$. For $0.020 \lesssim \varepsilon \lesssim 0.080$, with fixed parameters $\alpha$ and $\varepsilon$, the network can either be phase-synchronized, with $\langle R \rangle \approx 0.90$, or even be non-phase-synchronized, with $\langle R \rangle \approx 0.05$. The difference is simply due to different shuffling over the sequences of input values $K_{i}$. This region also depicts a great difference between the maximum and minimum values of $\langle \delta R \rangle$. Higher values of this quantifier occur with smaller values of $\langle R \rangle$, in which case the network is only phase synchronized at the level of groups, but not globally. Otherwise, smaller values of $\langle \delta R \rangle$ occur with higher $\langle R \rangle$, in which case the entire network reaches PS. This phenomenon, called synchronization malleability, is observed in the region where $\chi(R)$ has high values (Fig. \ref{fig:main_sync} (b)).

For $\alpha = 2.5$ (green lines with diamonds), the network is always non-phase-synchronized, with $\langle R \rangle < 0.4$. Furthermore, we note that some of the values of $\langle R \rangle$ for the coupled network are smaller than in the uncoupled case ($\varepsilon = 0.0$). In fact, in these cases the values of $\langle R \rangle$ are similar to the expected value for randomly distributed phases ($R_{\mathrm{random}} \sim 1/\sqrt{N} = 0.0436$, considering $N = 525$) \cite{arenas2008synchronization}. Besides, the increase of $\varepsilon$ makes the maxima and minima of $\langle \delta R \rangle$ also increase, indicating that the groups of the network may be phase synchronized, even though the network as a whole may not.

Panel (c) of Fig. \ref{fig:order_cases} depicts the Kuramoto order parameter $\langle R \rangle$ as a function of $\varepsilon$ for the non-shuffled case, which is the sequence of input values $K_{i}$ given by Eq. (\ref{eq:inputs}) and represented by the black dots in Fig. \ref{fig:inputs}. The red line (big diamonds) represents the case of $\alpha = 1.0$, where the increase of $\varepsilon$ makes the network transition from non-synchronized states to phase-synchronized ones. Comparing with the shuffled cases (panel (a)), the non-shuffled case requires higher coupling strength to phase synchronize and even then it does so at a lower degree of $\langle R \rangle$. Increasing the locality parameter to $\alpha = 1.8$ (purple line with hexagons), the network is always non-phase-synchronized, a drastically different scenario than the shuffled cases, where the network can depict PS. At last, for $\alpha = 2.5$ (brown line with crosses), the network also depicts small values of $\langle R \rangle$ and does not reach PS. We therefore see that the shuffling may facilitate the PS of the network or even hinder it. This limit case, represented by the non-shuffled values of $K_{i}$, is just an artificial construction used to illustrate the effects of the shuffling process.

In panel (d) of Fig. \ref{fig:order_cases}, the dependence of the network to initial conditions is studied. It depicts the $\langle R \rangle$ of the network for two representative shuffled input sequences $\{K_{i}\}$ shown in Fig. \ref{fig:inputs}, labeled as shuffling \#1 and \#2. In this case, $30$ different initial conditions for $x_{i}$ and $y_{i}$ are simulated and the maximum and minimum values of $\langle R \rangle$ are shown by the extremes in the filled area as a function of $\varepsilon$ with $\alpha = 1.8$. For $\varepsilon \lesssim 0.020$, the networks depicts similar values of $\langle R \rangle$. However, for $0.020 \lesssim \varepsilon \lesssim 0.053$, this is changed, as we observe different values of $\langle R \rangle$ for the two cases. In this region the network has synchronization malleability: with only a difference between the sequence of inputs $K_i$, the network can present $\langle R \rangle \approx 0.88$, for shuffling \#1, or even $\langle R \rangle \approx 0.03$, for shuffling \#2. For the region of high coupling strength, $\varepsilon \gtrsim 0.053$ the value of $\langle R \rangle$ is still different, but both networks depict phase-synchronized states. The results also show that the initial conditions of the systems do not seem to affect the dynamics of the network, since the filled area, representing variations due to different initializations, is only visible in a small region around $\varepsilon \approx 0.022$. Therefore, synchronization malleability is observed for different inputs $\{K_i\}$, but not for different initial conditions. We performed tests considering different values of $\sigma$ and network size $N$ and the synchronization malleability phenomenon can still be observed.

\subsection{Spatiotemporal patterns}
\begin{figure*}[th]
    \centering
    \includegraphics[width=0.95\textwidth]{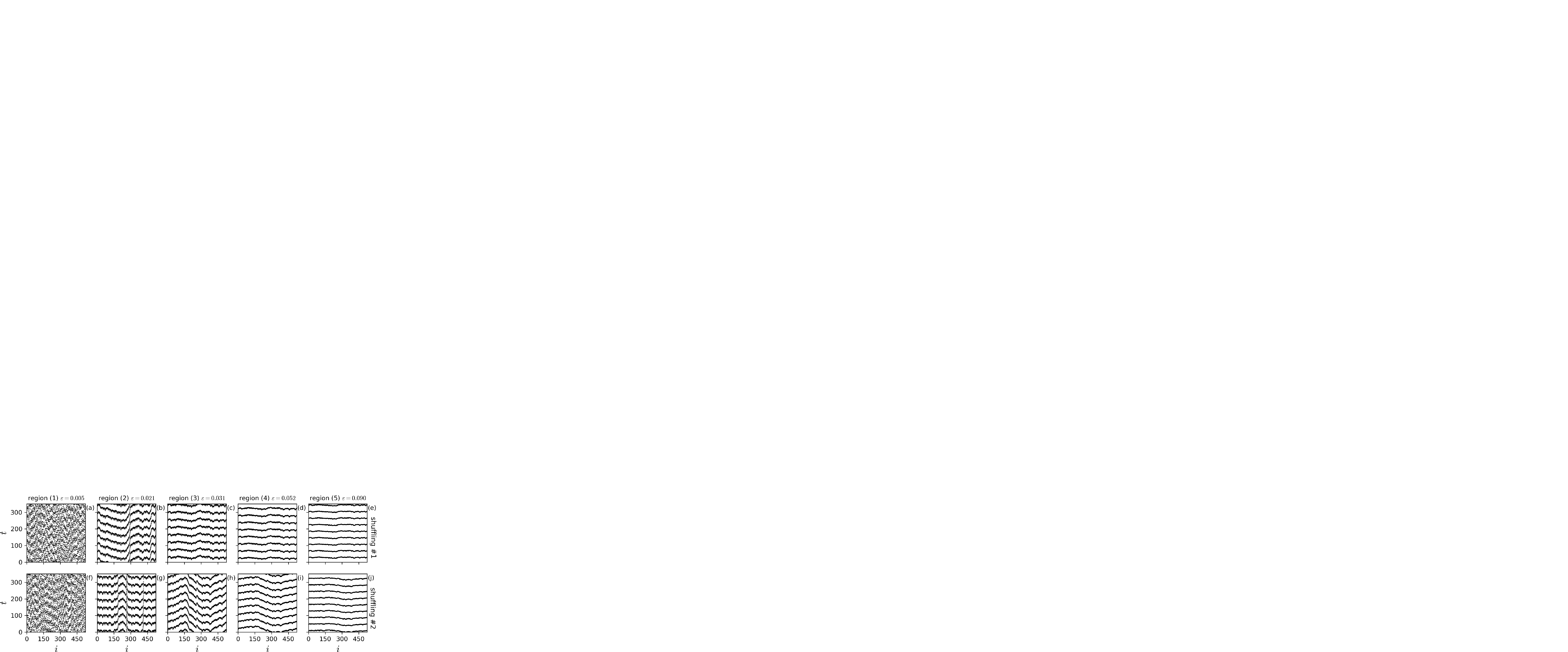}
    \caption{Raster plots obtained from the activation variable ($x$) show the spatiotemporal patterns of the network for different sets of parameters $\alpha$ and $\varepsilon$. Each region is indicated by the number in Fig. \ref{fig:main_sync} (a) and Fig. \ref{fig:order_cases} (d). The results depicted in the first row represent the network with the process of shuffling \#1, while the results in the second row represent the process of shuffling \#2. For all analyses here, $\alpha = 1.8$. Panels (a) and (f) depict non-synchronized states, where $\varepsilon = 0.005$; panels (b) and (g) ($\varepsilon = 0.021$) show diagonals structures; panels (c) and (h) ($\varepsilon = 0.031$) depict different dynamical states characterizing synchronization malleability. A similar scenario is observed in panels (d) and (i) ($\varepsilon = 0.052$) where the network with shuffling \#1 is phase synchronized ($\langle R \rangle = 0.88$) and the one with shuffling \#2 is not ($\langle R \rangle = 0.03$). At last, panels (e) and (j) show both cases with phase synchronization, with $\varepsilon = 0.090$.}
    \label{fig:raster_plot_1}
\end{figure*}

For a better visualization of the dynamical states exhibited by the network, Fig. \ref{fig:raster_plot_1} depicts raster plots (RP) of the spike times. Each black dot in the figure represents the time when a spike starts for the $i$-th neuron. We set $\alpha = 1.8$ for all cases and consider different values of $\varepsilon$ for shuffling \#1 (first row) and shuffling \#2 (second row). Regions indicated in the titles are the same ones defined in Fig. \ref{fig:main_sync} (a) and Fig. \ref{fig:order_cases} (d). Therefore, panels (a) and (f) are representative of non-synchronized states ($\varepsilon = 0.005$), where the spikes start at different times and there is no coherence in the neuronal activity. Panels (b) and (g) represent the case where $\varepsilon = 0.021$ and $\langle R \rangle$ is similar for both shuffles. In this case, one can observe diagonal structures and parts of the network with PS, which is in line with $\langle \delta R \rangle$ starting to depict high values (see Fig. \ref{fig:main_sync} (b)).

A further increase in the coupling strength leads the networks to different dynamical states: panels (c) and (h) represent the case for $\varepsilon = 0.031$, where the network with shuffling \#1 reaches PS ($\langle R \rangle = 0.79$) and horizontal structures can be observed, while the network with shuffling \#2 does not reach PS ($\langle R \rangle = 0.18$) and, instead, diagonal structures can be observed. Instead of network PS, shuffling \#2 has only groups of neurons with PS. For $\varepsilon = 0.052$, in panels (d) and (i), the dynamical difference between the networks increases: the network in (d) (shuffling \#1) is phase-synchronized ($\langle R \rangle  = 0.88$), but the one in (i) (shuffling \#2) is not ($\langle R \rangle = 0.03$). In the latter case, one can notice the existence of locally horizontal structures in the raster plots, showing the existence of phase-synchronized groups that are not phase synchronized among themselves. This leads to the entire network having a small value of $\langle R \rangle$. This situation is similar to the phenomenon of anti-phase synchronization, where groups of oscillators depict, separately, synchronized characteristics, which cancel themselves globally \cite{kamal2015emergent,xie2010synchronization}. At last, for the region of high coupling strength $\varepsilon = 0.090$, both networks depict phase-synchronized states and similar dynamical behaviors (panels (e) and (j)). The previous phenomena are observed for other shuffled input sequences and for other values of $\varepsilon$ where the synchronization malleability is detected (see Fig. \ref{fig:main_sync} (c)). 

The results characterize the effect of the coupling strength variation for a fixed value of $\alpha$. In this sense, as $\varepsilon$ increases, one can observe a transition from non-phase-synchronized states to states where diagonals structures and local-phase synchronization are observed to, finally, phase-synchronized states. This kind of transition is observed for all networks with different shuffling processes for $1.5 \lesssim \alpha \lesssim 2.0$, which is the region in the parameter space where $\langle \delta R \rangle$ has high values (see Fig. \ref{fig:main_sync} (c)).

\begin{figure}[htb]
    \centering
    \includegraphics[width=0.95\columnwidth]{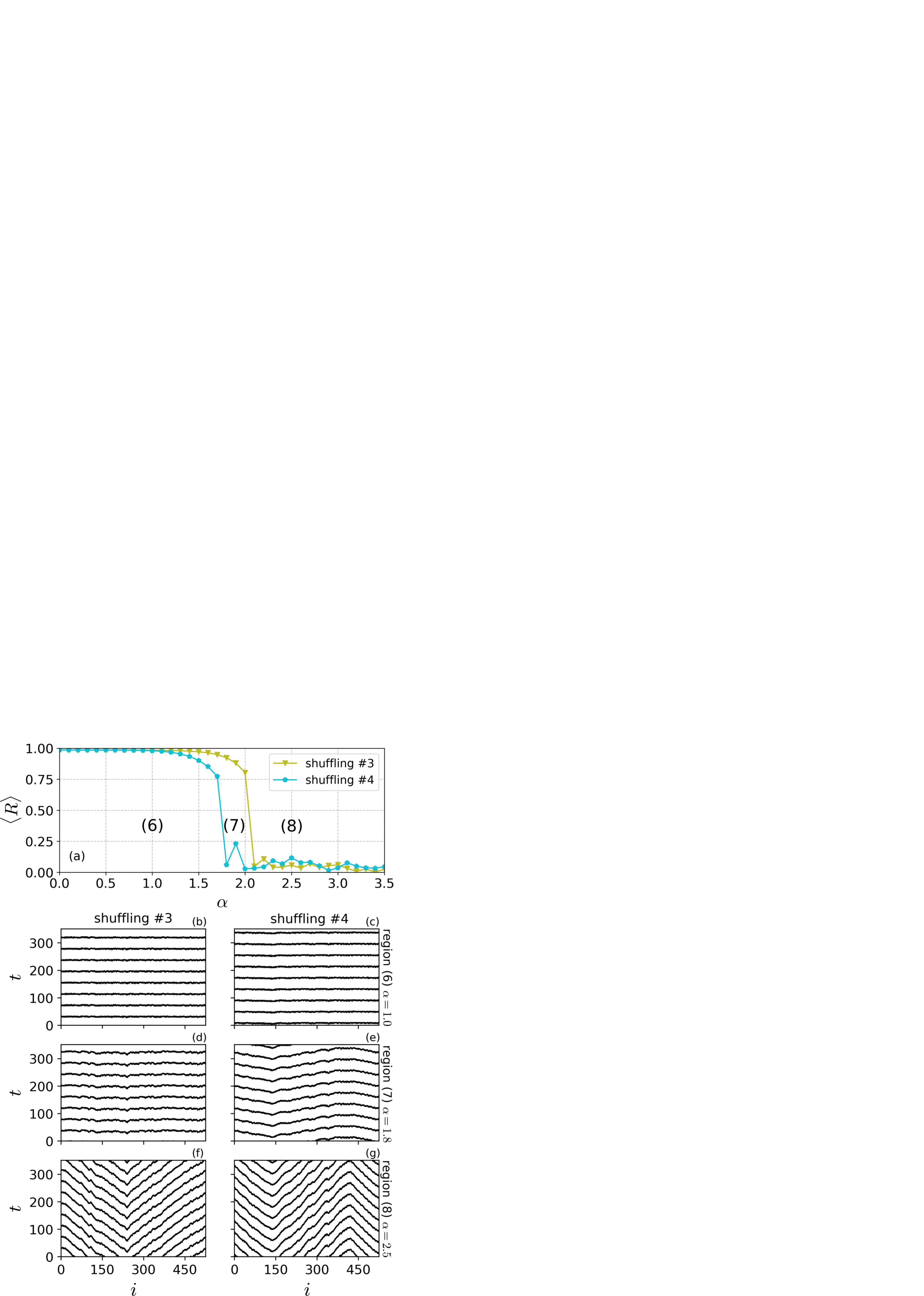}
    \caption{The influence of an increase in $\alpha$ is shown in panel (a), where $\langle R \rangle$ is depicted for $\varepsilon= 0.070$ considering two sequences $\{K_i\}$ of input values  given  by  the  shuffling  processes  \#3  (olive  line with down triangles) and \#4 (cyan line with pentagons).  The raster plots are obtained considering $\varepsilon = 0.070$ and different values of $\alpha$. Panels (b) and (c) depict the raster plots for phase-synchronized states ($\alpha = 1.0$); panels (d) and (e) show the case where the synchronization malleability is observed ($\alpha = 1.8$), and, at last, panels (f) and (g) ($\alpha = 2.5$)  are representative of the dynamical states where diagonal structures are noticed.}
    \label{fig:raster_plot_2}
\end{figure}

Figure \ref{fig:raster_plot_2} depicts similar analyses, but considering a fixed value of coupling strength at $\varepsilon = 0.070$ and varying the locality parameter $\alpha$. In this way, panel (a) shows $\langle R \rangle$ as a function of $\alpha$ for two illustrative sequences $\{K_{i}\}$ of input values (shuffling \#3 - olive line with down triangles - and shuffling \#4 - cyan line with pentagons). For $\alpha \lesssim 1.5$, both networks depict PS, with high values of $\langle R \rangle$. However, for $1.7 \lesssim \alpha \lesssim 2.1$, one can observe different degrees of PS, represented by the difference in the values of $\langle R \rangle$ between the two cases. The olive line depicts $\langle R \rangle \gtrsim 0.8$ while the cyan line shows a very small degree of PS ($\langle R \rangle \lesssim 0.25$), which is a clear example of the synchronization malleability phenomenon. Finally, for $\alpha \gtrsim 2.1$ both networks depict similar dynamical behaviors with small values of $\langle R \rangle$.

In order to analyze the spatiotemporal patterns, raster plots are plotted in the other panels of Fig. \ref{fig:raster_plot_2}. In this case, the input values $\{K_{i}\}$ are given by shuffling \#3 and \#4. The values of $\alpha$ correspond to the regions 6, 7, and 8 defined in Fig. \ref{fig:main_sync} (a) and Fig. \ref{fig:raster_plot_2} (a). The results shown in panels (b) and (c) ($\alpha = 1.0$) indicate that both networks are in a phase-synchronized state, corroborating the high values of $\langle R \rangle$ observed in panel (a) for this region. However, for $\alpha = 1.8$, the synchronization malleability phenomenon is again observed: panel (d) (shuffling \#3) depicts a phase-synchronized state, with $\langle R \rangle = 0.92$, while panel (e) (shuffling \#4) depicts does not, with $\langle R \rangle = 0.05$. At last, for higher values of $\alpha$, the networks depict similar dynamical states without PS. Panels (f) and (g) ($\alpha = 2.5$) show interesting spatiotemporal patterns with diagonal structures. This phenomenon can be understood as the effect of a stronger distance-dependence in the topology since higher values of $\alpha$ lead the effective coupling only with neighbors of each neuron. In this sense, coherence in the entire network is not obtained. Similar behavior is found in different models and coupling architecture when the coupling scheme has local characteristics \cite{osipov2005synchronized, wang2008synchronization, wang2010impact}.

\begin{figure}[htb]
    \centering
    \includegraphics[width=0.95\columnwidth]{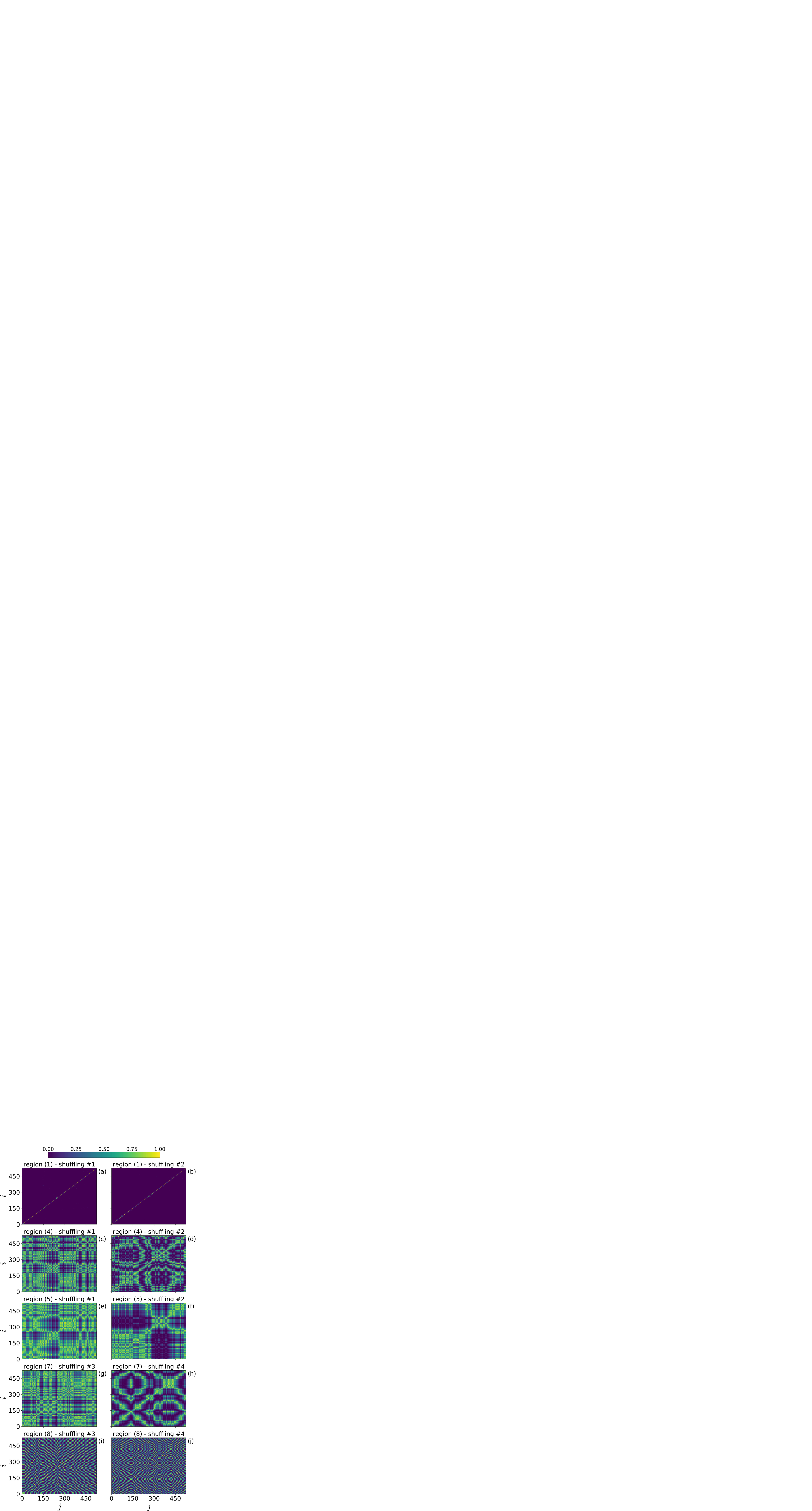}
    \caption{Heatmap of the functional connectivity, measured as the mutual information between neurons (see section \ref{sec:3.3}), showing different patterns of phase synchronization.  The title of each panel contains the region and shuffling number defined in previous figures and are as follows: panels (a) and (b): $\alpha = 1.8$, $\varepsilon = 0.005$; panels (c) and (d): $\alpha = 1.8$, $\varepsilon = 0.052$; panels (e) and (f): $\alpha = 1.8$, $\varepsilon = 0.090$; panels (g) and (h): $\alpha = 1.8$, $\varepsilon = 0.070$; panels (i) and (j): $\alpha = 2.5$, $\varepsilon = 0.070$.}
    \label{fig:mutual_information}
\end{figure}

%
To better understand why different shuffles of inputs $\{K_{i}\}$ lead to differences in the behavior of the network, we analyze its functional connectivity (FC). This is considered to be the matrix $(\mathbf{MI})_{XY} = MI(X,Y)$ of mutual information between neurons, as described in section \ref{sec:3.3}. Figure \ref{fig:mutual_information} shows the results of this procedure for the most noteworthy behaviors of the network. 

The main importance of the FC analysis is the identification of the synchronization characteristics of the network since mutual information between pairs of neurons calculated in this way is proportional to their degree of PS \cite{palus1997detecting}. In this context the results in Fig. \ref{fig:mutual_information} can be compared with the spatiotemporal pattern depicted by the raster plots of networks (Figs. \ref{fig:raster_plot_1} and \ref{fig:raster_plot_2}): the horizontal lines observed in the FC are related with horizontal structures in the RP. It is clear then that, in cases where the network reaches PS, the horizontal structures in the FC are bigger. This means that neurons share more information and are more phase synchronized at both local (neighborhood) and global (long-range) levels.

The cases depicted in panels (a) and (b), with $\varepsilon = 0.005$, are simple: the network is desynchronized (small value of $\langle R \rangle$), which shows in the FC as mutual information being non-zero only on the main diagonal. A different scenario is observed in panels (c) and (d), where $\varepsilon = 0.052$. In panel (c), with shuffling \#1, the FC is dominated by long horizontal and vertical structures (in yellow tones), indicating the sharing of information (PS) between both neighboring and distant neurons. Otherwise, for panel (d), with shuffling \#2, horizontal and vertical structures are shorter, indicating a lesser degree of PS between neurons and, therefore, smaller $\langle R \rangle$. In fact, the network in panel (c) has $\langle R \rangle = 0.88$, while the one in (d) has $\langle R \rangle = 0.05$. This is a case of synchronization malleability: some input sequences $\{K_i\}$ facilitate the formation of local and global structures, leading to PS, while others facilitate formation only of local structures, thus leading to some groups phase-synchronized within themselves, but not between themselves. 

A further increase of the coupling strength to $\varepsilon = 0.09$ (panels (e) and (f)) leads both networks (with shuffling \#1 and \#2) to PS. The structures observed in panels (c) and (d) are intensified in this case, with values of the mutual information increasing. Both FCs now share information at local and global levels, meaning both networks are phase synchronized. However, panel (e) (shuffling \#1) has longer horizontal and vertical structures than panel (f) (shuffling \#2), indicating a higher degree of PS, which is indeed the case: $\langle R \rangle = 0.95$ for (e) versus $\langle R \rangle = 0.82$ for (f) (see Fig. \ref{fig:order_cases} (d)).

To further illustrate the origin of the synchronization malleability, we show in panels (g) and (h) of Fig. \ref{fig:mutual_information} another example of different structures being formed due simply to a reordering of neurons' inputs ($K_{i}$). This is the extreme case depicted in Fig. \ref{fig:raster_plot_2} (d) and (e), where  $\langle R \rangle = 0.92$ and $\langle R \rangle = 0.05$, respectively. Again, the case of global phase synchronization happens due to the formation of local and global structures, while the other happens due to the dominance of local structures.

At last, panels (i) and (j) are representative of states with diagonal structures and zig-zag states, observed for higher values of $\alpha$ (see panels (f) and (g) of Fig. \ref{fig:raster_plot_2}). For both cases, higher values of mutual information between neurons are observed at diagonal lines in the FC, which correspond to the diagonal structures in the RP. In this case, the network is only frequency synchronized, but not phase synchronized. Furthermore, we note that, although $\langle R \rangle$ is similar to cases of very low coupling, like $\varepsilon = 0.005$ (for example, $\langle R \rangle \approx 0.07 $ in panel (i) versus $\langle R \rangle \approx 0.068$ in (a)), the dynamics of the network is different, with a formation of various structures in this case.  

\subsection{Hypothesis and mechanisms of malleability}

We have shown so far that different neural inputs can lead to largely different functional connectivities and synchronization properties of malleable networks. To investigate which properties of the inputs lead to the corresponding network behavior, we start with the hypothesis that neurons with similar features tend to synchronize in excitatory networks. Based on that, we can propose that shuffling processes over $K_i$ can cause, by chance, agglomeration of similar neurons across the network in some cases, making these networks more synchronized. To test this, we: (i) divide the network into groups of neurons by binning them (cf. section \ref{sec:3.1}); (ii) evaluate the firing rate (fr) of each neuron (each $K$ value leads to a different firing rate \cite{chialvo1995generic}) and then evaluate the mean firing rate of each bin, or group of neurons; (iii) we evaluate the standard deviation over all bins and obtain the coefficient of variability (CV) (standard deviation divided by the mean value of all bins); (iv) test for different numbers of neurons in each group (bin sizes).

\begin{figure}[htb]
    \centering
    \includegraphics[width=0.925\columnwidth]{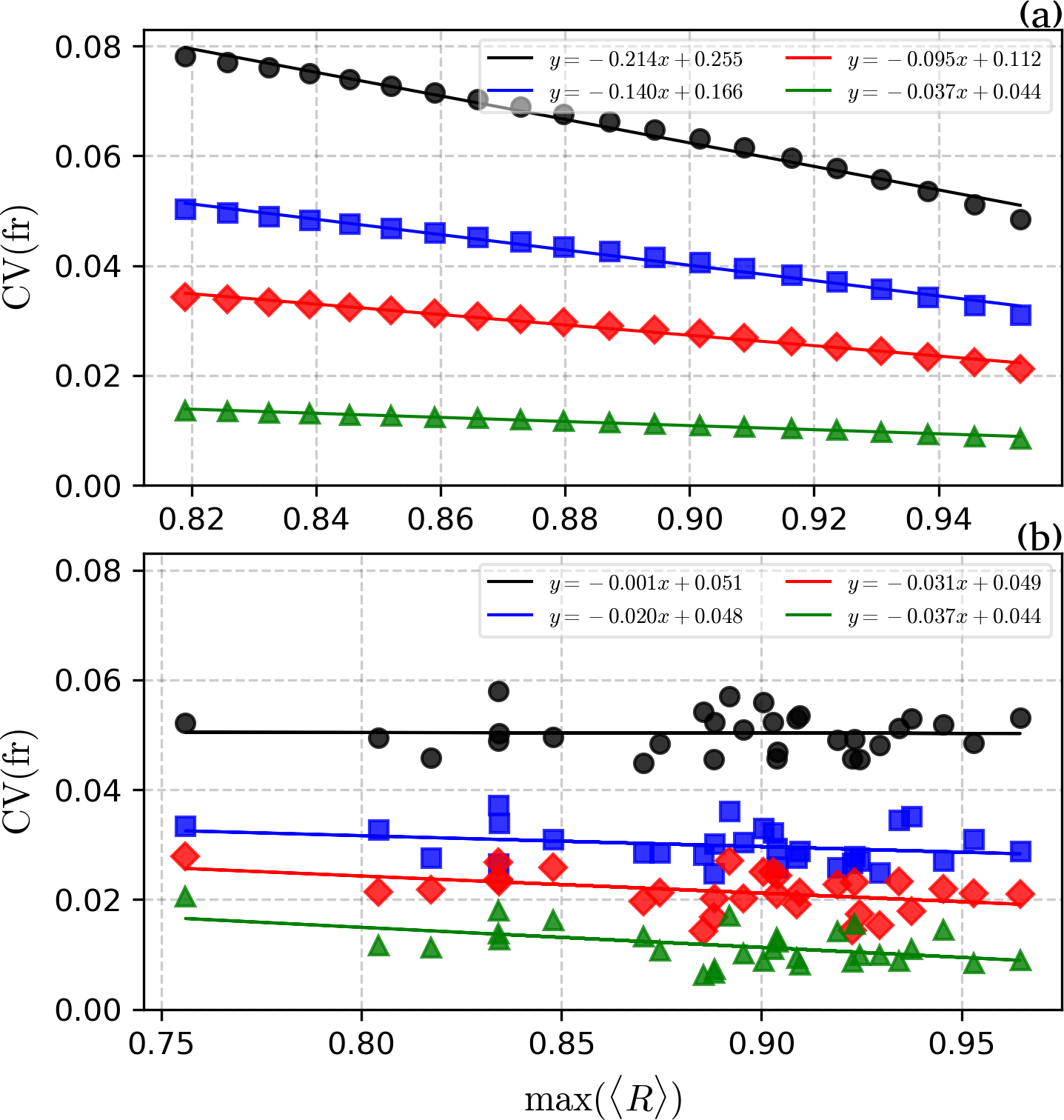}
    \caption{Coefficient of variability of neurons' firing rate (CV(fr)) as a function of $\mathrm{max}(\langle R \rangle)$. Panel (a) shows the case where the positions of $K$ values follow the shuffling \#1 and $\sigma$ is increased, in which the higher the CV(fr) the lower the level of synchronization. On the other hand, panel (b) shows a case of the malleable region, where $\sigma = 0.0035$, $\alpha = 1.80$, and different shuffles are considered and there is no clear relation between agglomeration of similar neurons and the collective behavior. The analyses are performed considering different bins sizes: $5$, $15$, $25$, and $75$ neurons in each one, which is represented by the black circles, blue squares, red diamonds, and green triangles, respectively.}
    \label{fig:variability_synchronization}
\end{figure}

First, by fixing the position of each neuron, the increase of the dissimilarity parameter $\sigma$ leads to networks with consecutively higher variability CV(fr), with a corresponding lower degree of synchronization. A representative case is depicted in Fig. \ref{fig:variability_synchronization}(a) considering the network with shuffling \#1 and $\alpha = 1.80$, where a linear relation is observed between CV(fr) and the maximum value of Kuramoto order parameter observed ($\mathrm{max}(\langle R \rangle)$ for each case. Moreover, the higher the CV(fr) the lower the level of synchronization, which indicates that relative positions have relevance for the network. Here, we consider different bins sizes: 5 neurons, 15 neurons, 25 neurons, and 75 neurons in each one, which is represented by the black circles, blue squares, red diamonds, and green triangles, respectively.

On the other hand, for the networks with malleability, a relation between CV(fr) and synchronization is not observed.  Figure \ref{fig:variability_synchronization} (b) shows the analysis for representative cases of malleability where $\alpha = 1.80$ and $30$ different shuffling processes over $K_{i}$ are considered (the same depicted in Fig. \ref{fig:order_cases}). Despite linear fits indicate positive slopes for data in (b), the relation between variability due to different shuffles and synchronization is not clear as in the case of the panel (a). The dispersion of data is very high and we cannot validate the hypothesis using these results. Besides this, we performed different tests considering directly $K_{i}$ values instead of firing rate, the relation between statistical properties of $K_{i}$ and synchronization, include different synchronization features than showed here, as the (critical) value of coupling strength when the Kuramoto order parameter increases above some threshold, or the integral of $\langle R \rangle(\varepsilon)$ but the results are very similar and we are not able to find a simple relation between the input sequence and the network behavior.

In fact, it is possible that no simple mechanism exists, and linear relations may be not able to explain malleability at all. To show this, we consider the networks with $\alpha = 1.80$ and shuffling \#1 and \#2 (the same used in Fig. \ref{fig:order_cases} (d)) and randomly choose one pair of neurons and switch their input values $K$. This procedure leads to a slightly different network, but there is no relevant difference in the agglomeration features and/or statistical properties of $K_{i}$. We then analyze the network synchronization for different pairs of neurons ($50$ simulations).

\begin{figure}[htb]
    \centering
    \includegraphics[width=0.9\columnwidth]{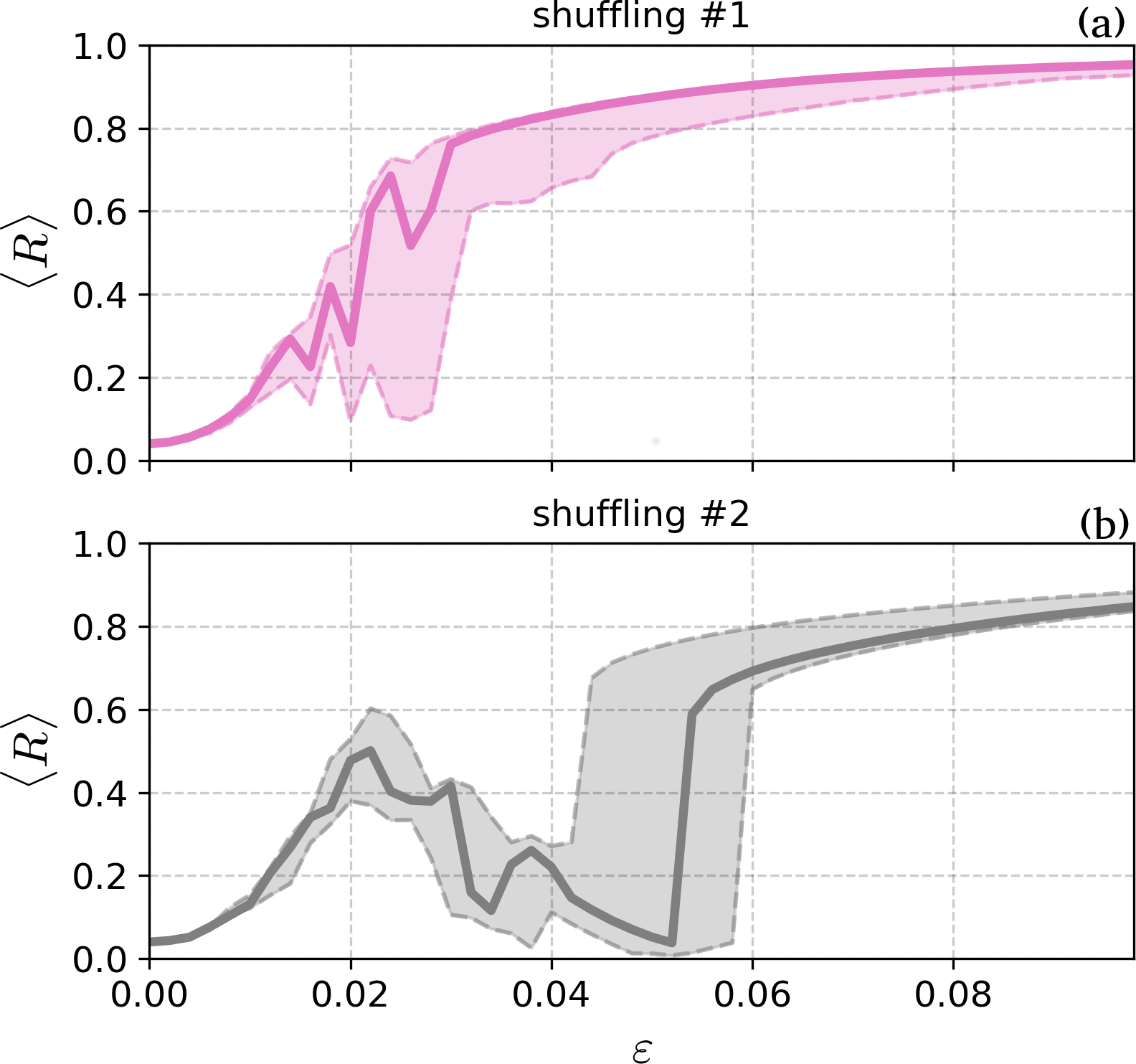}
    \caption{$\langle R \rangle$ is depicted as a function of $\varepsilon$ considering $50$ simulations where one pair of neurons has the $K$ value switched. Panel (a) represents the case of shuffling \#1 and panel (b) \#2. The extremes of filled areas indicate the maxima and minima values of $\langle R \rangle$ considering all simulations the thick lines represent the results without any switching. The results indicate that networks with very similar sets of $K_{i}$ values can have different synchronization properties.}
    \label{fig:max_min_one}
\end{figure}

The results are depicted in Fig. \ref{fig:max_min_one}, where panel (a) shows the case of shuffling \#1 and panel (b) of shuffling \#2. The thick line in each case represents the network without the pair switching, and the extremes of the filled area represent the maxima and minima values of $\langle R \rangle$ for each $\varepsilon$ considering the $50$ simulations. One single pair switch can considerably change the network behavior, either desynchronizing (as in panel (a)) or synchronizing ((b)) it. There was no clear relation between the properties of the switched pair (the difference between $K$s, firing rates, or indexes) and the changes in the network behavior. That is, even for the simplest case of a single switch, the behavior does not appear to be generated by a simple mechanism.

Given the results, we remark that in these networks similarity does facilitate synchronization, but in a more global fashion: networks with more similar neurons tend to synchronize more. This rule does not seem to apply locally, i.e. to a few neurons in a network. It appears that, especially in the malleable region, both short and long-range coupling are relevant, and the behavior can only be understood by considering the whole network.

\section{Conclusions}\label{sec:5}

Throughout this paper, we have analyzed a network composed of $525$ spiking neurons simulated with the Chialvo map following a connection architecture described by a distance-dependent power-law scheme. We have shown that the interplay between coupling strength $\varepsilon$ and power-law exponent $\alpha$ generates a great diversity of dynamical states. In networks with a weaker distance-dependence, a transition from desynchronized to phase synchronized states is observed with an increase in the coupling. By making the distance-dependence stronger, the network loses the phase-synchronized feature and there is a formation of new synchronization patterns, with locally phase-synchronized states and diagonal structures with only frequency synchronization. Similar behavior is found considering different local dynamics \cite{anteneodo2003analytical, batista2002lyapunov, li2006phase}.

We have found a region in the parameter space $\alpha \times \varepsilon$ where the networks depict a high level of sensitivity to changes in the neurons' inputs. In this case, very different dynamical states are observed for the same set of parameters ($\alpha$, $\varepsilon$): networks can be either phase-synchronized or phase-desynchronized, depending on the ordering of the inputs, but not on the initial condition. By calculating the mutual information between the pairs of neurons in the network, we have obtained its functional connectivity and characterized its patterns of phase synchronization. We have seen that, for the malleable region, the sequences of inputs can either facilitate or hinder the formation of phase-synchronized structures. For sequences facilitating global structures, the network reaches phase synchronization. Otherwise, it does not, and phase synchronization is reached only between smaller groups of neurons. 

Synchronization malleability was also found in a network of bursting neurons with coupling architecture following the Watts-Strogatz route, in which the network is taken from regular (local connections only) to small-world to random by the rewiring of connections \cite{budzinski2019synchronous}. In this work, the phenomenon is observed when some local connections are changed by non-local ones, generating a coexistence of local (neighborhood) and global (long-range) topological effects. We suggest that a similar case also occurs in our results, with malleability occurring when the locality parameter $\alpha$ has intermediate values and leads to networks with a mix of local and global effectiveness. 

Finally, this paper serves to characterize the phenomenon of synchronization malleability, for which there are still several open questions. We have shown that simple rules or linear relations between collective behavior and individual neural inputs do not seem to be enough to explain the mechanism behind malleability. We have demonstrated that very similar neural inputs can lead to different dynamical states, indicating malleability arises from a complex interplay between coupling, connection architecture, and individual neural inputs.

\section*{Acknowledgement}

This study was financed in part by the Coordena\c c\~ao de Aperfeiçoamento de Pessoal de N\'{\i}vel Superior - Brasil (CAPES) - Finance Code 001, Conselho Nacional de Desenvolvimento Cient\'{\i}fico e Tecnol\'ogico,  CNPq - Brazil, grant numbers 302785/2017-5 and 308621/2019-0, and Finan\-ciadora de Estudos e Projetos (FINEP). We thank the anonymous referees for contributing to a better manuscript.

\section*{Appendix}\label{appendix}

\begin{figure}[htb]
    \centering
    \includegraphics[width=0.95\columnwidth]{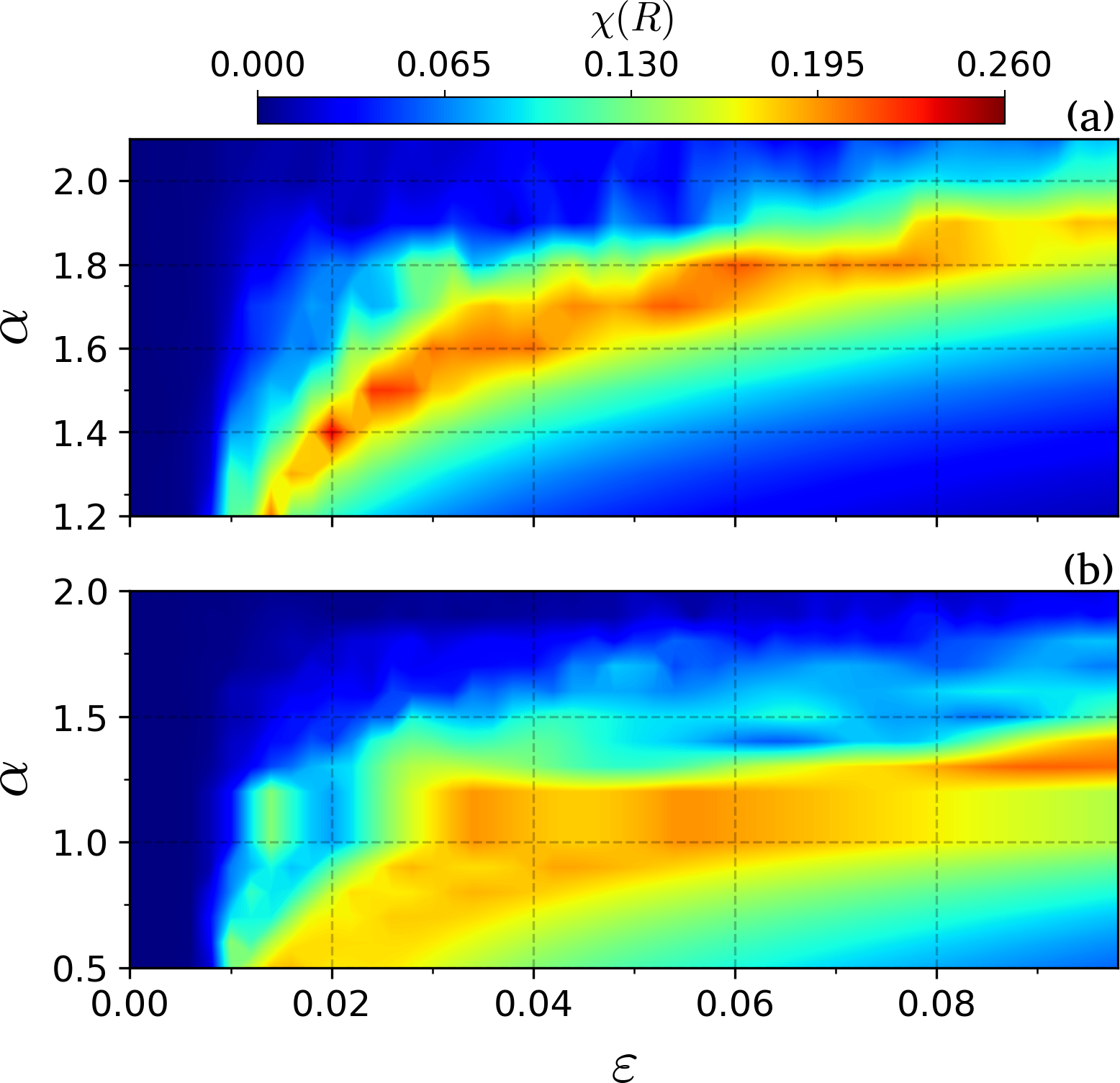}
    \caption{Dispersion of $\langle R \rangle$ over $30$ simulations considering different shuffles as a function of $\alpha$ and $\varepsilon$. In this case, higher values of the quantifier $\chi(R)$ indicate that the network assumes different levels of synchronization due to different shuffling in the input values. Here, we consider $N = 1001$ (a), and $N = 4001$ (b), where in all cases a malleable region is observed.}
    \label{fig:chi_sizes}
\end{figure}

We performed tests considering different network sizes and the malleability phenomenon remains. To exemplify this point, Fig. \ref{fig:chi_sizes} depicts the dispersion of $\langle R \rangle$ over $30$ simulations with different shuffling processes over $K_{i}$ values, which is represented by the quantifier $\chi(R)$, in the parameter space of $\alpha \times \varepsilon$. It is important to notice that higher values of this quantifier indicate that different shuffles lead to different levels of synchronization. Here, we consider the cases of $N = 1001$ (a), and $N = 4001$ (b), where there is a region in the parameter space of $\varepsilon \times \alpha$ where higher values of $\chi(R)$ are observed, indicating the existence of synchronization malleability. At last, other values of $N$ are considered and the conclusions remain.


%

\end{document}